\documentclass[fleqn,twoside]{article}
\usepackage{espcrc2}

\newcommand{\AmS}{{\protect\the\textfont2
  A\kern-.1667em\lower.5ex\hbox{M}\kern-.125emS}}

\title{Center Symmetry and Abelian Projection
at Finite Temperature}

\author{Michael C. Ogilvie\address[MCSD]{Washington University,
        St. Louis, MO, USA}%
        \thanks{We gratefully acknowledge the support of the U.S. Dept. of
                Energy under DOE DE-FG02-91ER40628.}
      } 
\begin{document}

\begin{abstract}
At finite temperature, there is an apparent conflict between Abelian
projection and critical universality. For example, should the deconfinement
transition of an $SU(2)$ gauge theory projected to $U(1)$ lie in the $Z(2)$
universality class of the parent $SU(2)$ theory or in the $U(1)$ universality
class? I prove that the projected theory lies in the universality class of
the parent gauge theory. The mechanism is shown to be non-local terms in the
projected effective action involving Polyakov loops. I connect this to the
recent work by Dunne \textit{et al.}~on the deconfinement 
transition in the 2+1
dimensional Georgi-Glashow model.
\vspace{1pc}
\end{abstract}

\maketitle

There is an apparent conflict between Abelian projection and critical
universality. In its simplest form, critical universality states that the
universality class of a phase transition depends only on the dimensionality
of the system and the symmetry group of the order parameter. Universality
thus tells us that the nature of the deconfinement transition for a pure
gauge theory with gauge group G depends only on the dimensionality of space
and the center of the gauge group C(G)\cite{Yaffe:qf}.

For example, $SU(2)\,$gauge theories are in the $Z(2)\,$%
universality class of the Ising model, while $SU(3)$ gauge theories are in
Potts model, or $Z(3)$, universality class. In Abelian projection, the gauge
group is reduced. This reduction can changes the center of the gauge group,
and thus naively may change the universality class. A common example would
be the reduction of $SU(2)$ to $U(1)$. The center of $SU(2)\,$is $Z(2)$, but
the center of $U(1)\,$is $U(1)\,$itself. Thus the deconfinement transition
of the projected theory appears to be in a different universality class from
the original underlying gauge theory. 
However, this naive expectation is wrong: lattice simulations show that
the critical exponents for $SU(2)$ projected to $U(1)$
are identical to the critical exponents of the underlying $SU(2)$ gauge
theory \cite{Ejiri:1996sh}.
The explanation for this behavior is given below.

In lattice gauge theory, Abelian projection has a natural algorithmic
formulation. For analytical purposes, 
I will use the formalism developed
in \cite{Ogilvie:1998wu}, which
uses gauge fields $\ u_{\mu }(x)\in G$, site-based gauge-fixing fields $%
g(x)\in G$, andAbelian gauge fields $h_{\mu }(x)\in H\subset G$. Three
different actions play a role: the gauge action 
\begin{equation}
S_{g}=\frac{\beta }{2N}\sum_{plaq}\,Tr\,\left( u_{p}+u_{p}^{+}\right),
\end{equation}
the gauge-fixing action, which for $SU(2)$ is given by 
\begin{equation}
S_{gf}=\lambda \sum_{links}Tr\,\left[ u_{l}\,\sigma _{3}u_{l}^{+}\sigma
_{3}\right], 
\end{equation}
and the projection action 
\begin{equation}
S_{proj}\left[ u,h\right] =\sum_{links}\left[ \frac{p}{2N}Tr\left(
u_{l}^{+}h_{l}+u_{l}^{+}g_{l}\right) \right] 
\end{equation}
As $p,\lambda \rightarrow \infty $, the usual lattice projection algorithm
is formally obtained. The variables $g\,$and $h$ are quenched: $u$'s are
generated by $S_{g}$, then $g$'s are generated by $S_{gf}$. Finally, $h$'s
are generated from the gauge-fixed $u$'s. 
Expectation values are given by
\begin{eqnarray}
\left\langle O\right\rangle &=&\frac{1}{Z_{g}}\int \left[ du\right]
\,e^{S_{g}\left[ u\right] }\frac{1}{Z_{gf}\lbrack u\rbrack }\int \left[
dg\right] e^{S_{gf}\left[ \widetilde{u}\right] }\nonumber \\
&&\cdot \frac{1}{Z_{proj}\lbrack 
\widetilde{u}\rbrack }\int \left[ dh\right] e^{S_{proj}\left[ \widetilde{u}%
,h\right] }\,O. 
\end{eqnarray}

Relations between the Green functions of the projected theory and the Green
functions of the underlying non-Abelian gauge theory can be found using
character expansions. The projection weight can be expanded as
\begin{eqnarray}
\lefteqn{ \exp \left[ \frac{p}{2N}Tr\left( u^{+}h+u^{+}g\right) \right] 
\nonumber }\\
&&=\sum_{\alpha
}d_{\alpha }c_{\alpha }\left( p\right) \chi _{\alpha }\left( h^{+}u\right) 
\end{eqnarray}
where $\chi _{\alpha }$ is a group character, $d_{\alpha }$ is the
dimensionality of the representation, and $c_{a}\left( p\right) $ is a
positive, p-dependent coefficient. I use $\widetilde{\chi }^{\beta }(h)$ to
denote the corresponding character of the projection subgroup. Note that $%
\chi ^{\alpha }(h)$ is in general a reducible representation of H. For
example, the j=1 representation of SU(2) is the sum of three irreducible
representations of U(1), given by m=+1,0,-1. To leading order in the
character expansion for projection, and to order $\lambda ^{0}$ for
gauge-fixing, we have
\begin{eqnarray}
\lefteqn{
\left\langle \widetilde{\chi }^{\beta }(h_{1}..h_{n})\right\rangle
\nonumber } \\
&&=\sum_{\alpha }\left( \frac{c_{\alpha }(p)}{c_{0}(p)}\right)
^{n}\int_{H}(dh)\,\widetilde{\chi }^{\beta }(h)\chi ^{\alpha
}(h^{+}) \nonumber \\
&& \ \ \ \cdot \left\langle \chi ^{\alpha }(g_{1}..g_{n})\right\rangle
\end{eqnarray}  
for any closed Wilson loop, where $n\,$is the length of the loop. For a
Polyakov loop $u_{P}(\overrightarrow{x})$, we have 
\begin{equation}
\left\langle Tr\,\,h_{P}(\overrightarrow{x})\right\rangle \simeq \left( 
\frac{c_{F}(p)}{c_{0}(p)}\right) ^{n_{t}}\left\langle Tr\,\,u_{P}(%
\overrightarrow{x})\right\rangle 
\end{equation}
Just above a second-order critical point, we have 
\begin{equation}
\left\langle Tr\,\,u_{P}(\overrightarrow{x})\right\rangle \sim \left(
T-T_{c}\right) ^{\beta } 
\end{equation}
so we must have the same critical index $\beta $ in the projected theory as
in the original. Similar results hold for higher correlation functions and
thus for all critical indices. We conclude that the critical behavior of the
projected theory must be that of the original gauge theory.

We can understand this result on the basis of center symmetry.
A center symmetry transformation on the original gauge theory acts on all
the timelike links on a given time slice as 
\begin{equation}
u_{0}\left( \overrightarrow{x},t\right) \rightarrow z\,u_{0}\left( 
\overrightarrow{x},t\right) 
\end{equation}
where $z\,$is an element of the center C(G) of G. $S_{g}\left[ u\right] \,$%
is invariant under this transformation but the Polyakov loop $u_{P}\left( 
\overrightarrow{x}\right) $ is not. The deconfinement transition is the
spontaneous breakdown of center symmetry at high temperature. REF

We can define an effective action $S_{eff}\left[ h\right] $\ for
the projected theory as 
\begin{eqnarray}
e^{S_{eff}\left[ h\right] } &=& \frac{1}{Z_{g}}\int \left[ du\right]
\,e^{S_{g}\left[ u\right] 
+S_{gf}\left[ u\right] +S_{proj}\left[ u,h\right] }\nonumber \\
&&\cdot \frac{1}{Z_{gf}\lbrack u\rbrack }\frac{1}{Z_{proj}\lbrack u\rbrack } 
\end{eqnarray}
Now suppose that w is an element of C(H). Then consider the transformation $%
h_{0}\left( \overrightarrow{x},t\right) \rightarrow w\,h_{0}\left( 
\overrightarrow{x},t\right) $ on a single time slice. 
It is easy to see that the replacement
$S_{eff}\left[ h\right] \neq S_{eff}\left[ wh\right]$
unless $S_{g}\left[ wu\right] = S_{g}\left[ u\right]$.
Thus $S_{eff}$ is only invariant under $C(G)$, not $C(H)$.

The effective action can be constructed order by order in a strong-coupling
expansion around $\beta ,\lambda ,p=0$. 
To leading order for SU(N) we have
\begin{equation}
S_{eff}\left[ h\right] =\frac{\beta p^{4}}{32N^{9}}\sum_{plaq}\,Tr\,\left(
h_{plaq}+h_{plaq}^{+}\right) 
\end{equation}
This effective action is invariant under gauge transformations in $H$ and \
invariant under the global symmetry $h\rightarrow wh$. Note that
contributions from Wilson loops of area $A$ are suppressed by a factor of $%
\beta ^{A}$, but such terms are also invariant under $h\rightarrow wh$.

Polyakov loops break the invariance of $S_{eff}\left[ h\right] $ under $%
h\rightarrow wh$. The lowest such contribution comes from integration over $%
N_{t}$ temporal plaquettes, aligned to form a belt of plaquettes around the
lattice in the temporal direction. Integration over the $u$ field leads to
an interaction of the form 
\begin{eqnarray}
&&\left( \frac{\beta }{2N^{2}}\right) ^{N_{t}}
\left( \frac{p}{2N^{2}}\right)
^{2N_{t}}\nonumber \\
&&\cdot \left[ Tr\left( h_{P}(\overrightarrow{x})\right) Tr\left( h_{P}^{+}(%
\overrightarrow{x}+\widehat{i})\right) +h.c.\right] 
\end{eqnarray}
where $h_{P}(\overrightarrow{x})$ is a Polyakov loop formed from the $h$
fields: 
\begin{equation}
h_{P}(\overrightarrow{x})=\prod_{t}h_{0}\left( \overrightarrow{x},t\right). 
\end{equation}
This Polyakov loop term is not invariant under the 
change of variables $h_P\rightarrow wh_P$ unless $w$ is in $C(G)$.

Consider the case of $SU(2)$ projected to $U(1)$. If we write 
\begin{equation}
h_{P}(\overrightarrow{x})=\left( 
\begin{array}{ll}
e^{i\theta (\overrightarrow{x})} & 0 \\ 
0 & e^{-i\theta (\overrightarrow{x})}
\end{array}
\right),
\end{equation}
the interaction between Polyakov loops has the form 
\begin{equation}
\cos \theta (\overrightarrow{x})\cos \theta (\overrightarrow{x}+\widehat{i})%
.
\end{equation}
The transformation $h_P\rightarrow wh_P$ induces a change 
$\theta \rightarrow \theta + \alpha$
which is a symmetry of the action only when $\alpha =0$ or $\alpha =\pi $, 
\textit{i.e.}, when $w\in Z(2)$. A true $U(1)\,$gauge theory, in comparison,
could only have effective interactions like 
\begin{equation}
2\cos \left[ \theta (\overrightarrow{x})-\theta (\overrightarrow{x}+\widehat{%
i})\right] 
\end{equation}
which is invariant for arbitrary $\alpha $. 
We thus see how $S_{eff}\left[
h\right] $ includes Polyakov loop terms, non-local in Euclidean time, which
force the global center symmetry to be $C(G)$ rather $C(H)$. 

The need to include specific, non-local terms associated with
Polyakov loops to effective actions is not limited to the case of Abelian
projection. Similar behavior has recently been demonstrated by 
Dunne \textit{et al.}~\cite{Dunne:2000vp}
in the finite-temperature Georgi-Glashow model in $2+1$
dimensions. Adjoint representation scalars break an $SU(2)$ gauge symmetry
to $U(1)$. The low-energy behavior of the model is that of a $U(1)$ theory,
and thus the question of $Z(2)$ versus $U(1)$ center symmetry arises here as
well.

This model has a semiclassical regime where confinement can be demonstrated
as a consequence of instanton effects. 
Agasian and Zarembo \cite{Agasian:1997wv}
have shown that
at finite temperature, these instantons are seen by Polyakov loops as
vortices, of the type that drive Kosterlitz-Thouless phase transitions in $2$%
-dimensional systems. A detailed renormalization group analysis appeared
to show that the deconfining phase transition
in the $d=2+1$ Georgi-Glashow model was in the $U(1)$ universality class.

However, Dunne \textit{et al.}~have demonstrated that Polyakov loop
effects associated with heavy $W^{\pm }$ gauge bosons must be explicitly
included in any discussion of critical behavior. The effect on the
finite-temperature, $d=2+1$ Georgi-Glashow model is precisely parallel to
adding a symmetry-breaking term to a $U(1)$ spin model, reducing the global
symmetry from $U(1)$ to $Z(2)$. 
It is possible to show explicitly that the critical behavior of the model
lies in the $Z(2)$ universality class of the two-dimensional Ising model
after all.

Abelian projection does not enlarge center symmetry. As a
consequence, projection does not violate universality: the projected theory
is in the same universality class as the original, underlying gauge theory.
Any effective action for a projected theory must include non-local terms at
non-zero temperatures to give the correct critical behavior at
deconfinement. Comparison of projection at finite temperature with the
Georgi-Glashow model at finite temperature shows the similarity of the issue
in the two models, and the similar resolution. 
This suggests that we may have identified 
all the terms of the projected effective action which
can be relevant in the renormalization group sense.
If true, this is an important step in understanding how
Abelian projection reproduces the behavior of the underlying
gauge theory.

\end{document}